\documentclass[global,twocolumn]{svjour}

\usepackage{graphics}

\journalname{Applied Physics B}

\begin{document}

\title{Mid-infrared upconversion spectroscopy based on a Yb:fiber femtosecond laser}

\author{T. A. Johnson\thanks{\emph{Present contact information: }todd.johnson@csbsju.edu\newline Saint John's University, Collegeville, MN 56321 USA\newline phone: 320-363-3184, fax: 320-363-3202 } \and S. A. Diddams\thanks{scott.diddams@nist.gov}}                     
\institute{National Institute of Standards and Technology, Boulder, CO 80305 USA}
\date{Received: date / Revised version: date}
\maketitle

\begin{abstract}

We present a system for molecular spectroscopy using a broadband mid-infrared laser with near infrared detection.  Difference frequency generation of a Yb:fiber femtosecond laser produced a mid-infrared (MIR) source tunable from $2100-3700~{\rm cm^{-1}}$ $(2.7-4.7~{\rm\mu m})$  with average power up to 40~mW.  The MIR spectrum was upconverted to near-infrared wavelengths for broadband detection using a two-dimensional dispersion imaging technique.  Absorption measurements were performed over bandwidths of $240~{\rm cm^{-1}}$ (7.2~THz) with $0.048~{\rm cm^{-1}}$ (1.4~GHz) resolution, and absolute frequency scale uncertainty was better than $0.005~{\rm cm^{-1}}$ (150~MHz).  The minimum detectable absorption coefficient per spectral element was determined to be $4.4\times10^{-7}~{\rm cm^{-1}}$ from measurements in low pressure $\rm CH_4$, leading to a detection limit of 2 parts-per-billion.  The spectral range, resolution, and frequency accuracy of this system show promise for determination of trace concentrations in gas mixtures containing both narrow and broad overlapping spectral features, and we demonstrate this in measurements of air and solvent samples.

\end{abstract}

%%%%%%%%%%%%%%%%%%%%%%%%%%%%%%%%%%%%%%%%%%%%%%%%%%%
\section{Introduction}

Molecular spectroscopy has long been a useful tool in chemistry and physics.  As applications for molecular spectroscopy continue to grow into environmental sensing \cite{richter09}, security \cite{bauer08}, and medical analysis \cite{thorpe08}, the requirements for broad spectral range combined with good sensitivity become more demanding.  Traditional molecular spectroscopy has been accomplished with thermal sources in moving mirror Fourier transform spectrometers or with tunable single frequency lasers, but recent developments in femtosecond laser frequency combs combine the high brightness and frequency precision of a tunable laser with the broad spectrum previously open only to thermal sources.  

Effective molecular spectroscopy requires both the source at a useful wavelength range as well as a detection method to utilize the unique advantages of the source.  Broadband laser sources spanning $1.0-1.8~{\rm\mu m}$ wavelengths are readily available but  only able to observe weaker overtones of the strong hydrogen-stretch absorption features of many target molecules.  Sources in the mid-infrared wavelengths from $2.0-3.6~{\rm\mu m}$ would access much stronger fundamental absorption to allow more sensitive detection of these molecules.   Broadband femtosecond MIR lasers have been demonstrated using difference frequency generation between a comb and CW laser \cite{maddaloni06}, optical parametric oscillation \cite{adler09,leindecker11}, as well as direct comb generation \cite{sorokin07}.

Detection techniques that take advantage of the wide spectral bandwidth of a femtosecond source include using the laser as a high brightness source for a moving mirror Fourier transform spectrometer \cite{sorokin07,adler10} or dual frequency comb Fourier transform spectroscopy \cite{coddington08,bernhardt10,bernhardt10b}.  Dispersive methods for imaging a frequency comb spectrum have been shown with Vernier coupling of comb and cavity \cite{gohle07} or a high resolution etalon \cite{thorpe08,diddams07,cossel10}. Matching the mode spacing of a cavity to the repetition rate of a mode-locked laser allows broadband cavity enhanced absorption \cite{gherman04} or cavity ringdown spectroscopy \cite{thorpe06}.

We demonstrate a femtosecond laser system that generates broad spectra covering  an important mid-infrared absorption range of $2.7-4.7~{\rm\mu m}$  ($2100-3700~{\rm cm^{-1}}$) while retaining advantages of detection in the near infrared through an upconversion process.  A single measurement with up to 5000 elements at $0.048~{\rm cm^{-1}} (1.4~{\rm GHz})$ resolution enables simultaneous measurement of sharp spectral lines and broad absorption features from complex molecules.

%%%%%%%%%%%%%%%%%%%%%%%%%%%%%%%%%%%%%%%%%%%%%%%%%%%
\section{Experimental Details}
\subsection{Broad bandwidth mid-infrared source}

\begin{figure}
\begin{center}
\resizebox{\columnwidth}{!}{ \includegraphics{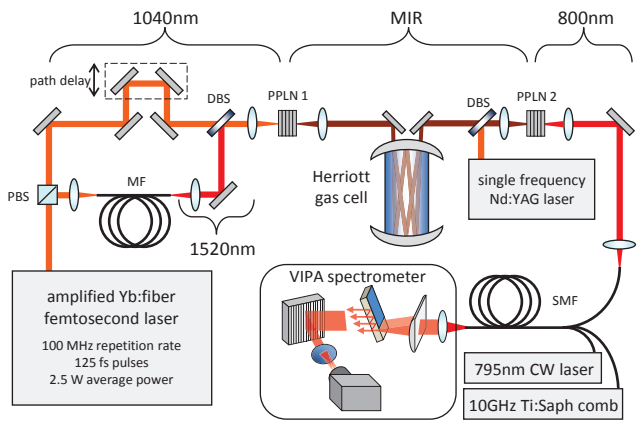}}
\end{center}
\caption{A Yb:fiber femtosecond laser was separated, broadened, and recombined on a nonlinear crystal to produce broad bandwidth mid-infrared spectra via difference frequency generation.  The mid-infrared beam was passed through a gas cell and upconverted on a second nonlinear crystal, then transferred to the two-dimensional spectrometer for analysis.  PBS=polarizing beam splitter, DBS=dichroic beamsplitter, MF=microstructured fiber, SMF=single mode fiber, PPLN=periodically poled lithium niobate crystal.}
\label{fig.setup}       
\end{figure}

\begin{figure}
\begin{center}
\resizebox{.8\columnwidth}{!}{ \includegraphics{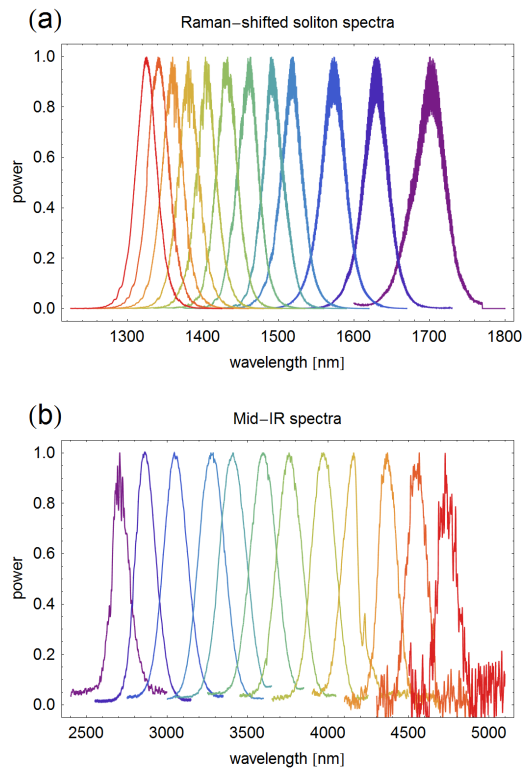}}
\end{center}
\caption{(a) Optical spectrum analyzer measurements of the filtered Raman solitons for average input powers 100-500~mW (average output power 10-20~mW).  Each spectrum normalized to peak value.
(b) Monochromator and InSb photodetector measurements of the MIR source.  Each spectrum normalized to peak value, with some spectra displaying absorption signals due to atmosphere (such as the carbon dioxide absorption near $4.2~{\rm \mu m}$).   Color of each signal spectrum in (a) corresponds to color of generated MIR spectrum in (b).
}
\label{fig.sigidler}       
\end{figure}

A Yb:fiber femtosecond laser was the source for a difference frequency generation (DFG) approach to a tunable, broad bandwidth MIR source.   The experimental setup is shown in Figure~\ref{fig.setup}.  Full details and characterization of this source are provided in a separate publication \cite{neely11}, while here we provide a brief summary of the properties most relevant for our spectroscopic measurements.  The MIR DFG portion of the system has similarities to Er:fiber comb approaches \cite{erny07,gambetta08,winters10}.  The output of an amplified Yb:fiber femtosecond laser provided 125~fs pulses at 100 MHz and 2.5~W average power.  This source was split into pump and signal paths, seeding a fan-out, periodically-poled ${\rm MgO:LiNbO_3}$  (PPLN) nonlinear crystal in a single pass to generate an idler wavelength ($\lambda_{\rm p}^{-1}-\lambda_{\rm s}^{-1}=\lambda_{\rm i}^{-1}$).  The pump path delivered 2.0-2.4~W average power to the PPLN after passing through a variable path delay stage.  The signal path coupled 100-500~mW average power into a 1~m length microstructured fiber to produce a Raman-shifted soliton that was tunable from $\rm 1.3-1.7~\mu m$ by varying coupled power (see Figure~\ref{fig.sigidler}a).  The pump and signal were spatially and temporally overlapped in the PPLN to produce the MIR light.  Figure~\ref{fig.sigidler}b shows the broad tuning possible by adjusting the in-coupled power to the microstructured fiber and the PPLN phase matching position.  After the PPLN, the MIR was separated from the pump and signal beams using a broadband anti-reflection coated germanium filter.   MIR generation efficiency varied with pulse parameters and crystal wavelength properties, and at the time of this work maximum power up to 40~mW occured near $\lambda_{\rm i}\approx3.3~{\rm \mu m}$. Observations reported in reference \cite{neely11} indicate coherence degradation of the Raman-shifted soliton and MIR light, however, the source had relatively low intensity noise ($<1\%$, 10 Hz - 10 MHz), a broad spectrum, and a well-collimated beam that were necessary for this work.

%%%%%%%%%%%%%%%%%%%%%%%%%%%%%%%%%%%%%%%%%%%%%%%%%%%%%%%%%
\subsection{Absorption path and gas samples}

The MIR beam was mode-matched into a multi-pass astigmatic Herriott cell (Figure~\ref{fig.setup}).  The total path length was $210~{\rm m}$, and mirror losses on the 238 reflections resulted in an optical transmission of 2\% for the cell.  A gas handling system was attached to the cell to prepare different samples, with total cell and system volume of 6~L.    Spectral artifacts (due to mirror or coating spectral dependence in the beam path) were minimized by comparing the measured gas absorption features with a normalization measurement of the evacuated cell.

%%%%%%%%%%%%%%%%%%%%%%%%%%%%%%%%%%%%%%%%%%%%%%%%%%%%%%%
\subsection{Virtually Imaged Phase Array (VIPA) and upconversion detection}

\begin{figure}
\resizebox{\columnwidth}{!}{ \includegraphics{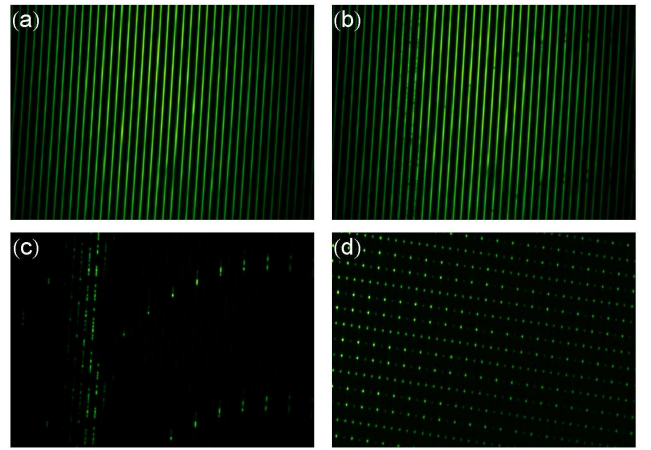}}
\caption{Two-dimensional VIPA spectrometer images captured by CCD.  (a) Dispersed source spectrum with no gas sample in the cell. (b) Methane absorption features imprinted on source spectrum. (c) Subtracted image (a)-(b), to highlight the absorption spectrum pattern on the VIPA image.  The VIPA's free spectral range is apparent in the vertical repetition of the absorption feature pattern at the top and bottom of image.   (d) 10~GHz Ti:Saph comb calibration grid, optical filter etalon causes variations in comb mode intensity.  }
\label{fig.vipa}       
\end{figure}

Rapid detection of broadband absorption signals was accomplished with a two-dimensional dispersion method using a Virtually Imaged Phase Array (VIPA) crossed with a conventional diffraction grating \cite{shirasaki96,xiao04}.  The VIPA acted as a highly dispersive transmission grating with 1.4~GHz ($0.048~{\rm cm^{-1}}$) resolution and free spectral range of order 100~GHz ($3.3~{\rm cm^{-1}}$).  The overlapping orders from the VIPA were spatially separated by the crossed grating and recorded on a CCD camera (Figures~\ref{fig.setup}~and~\ref{fig.vipa}a) to analyze a large source bandwidth simultaneously.  Spectroscopic measurements with a VIPA have been demonstrated in the visible \cite{diddams07} and near-infrared \cite{cossel10} wavelengths, but extension of the VIPA approach directly to the MIR has additional complications including optical material manufacturing limitations and the availability and sensitivity of cameras.

To enable use of the existing near infrared VIPA and cooled silicon CCD camera for MIR spectroscopy, our system utilized an additional upconversion step to transfer the longer wavelength spectral absorption features to a detectable wavelength range \cite{heilweil89,kubarych05}.  The MIR beam exited the gas cell with absorption features imprinted on the source spectrum and was combined on a dichroic beamsplitter with a CW 500~mW single frequency Nd:YAG laser at 1064nm (Figure~\ref{fig.setup}).  Both beams were focused into a second PPLN crystal phase-matched for sum frequency generation, producing upconverted pulses near 800nm.   The upconverted signal was transferred in single mode fiber to the VIPA optics.   The length of the upconversion PPLN was chosen to maximize the available MIR  and VIPA spectral ranges rather than total power, resulting in an average upconversion quantum efficiency of about $2\times10^{-5}$.   Combining the gas cell mirror losses and upconversion efficiency with the additional near-IR optical losses in the transfer fiber and VIPA spectrometer, the silicon CCD camera recorded about one photoelectron per $10^9$ MIR photons generated at the first PPLN crystal.  Despite this significant loss in signal, it was still possible to recover useful spectral information due to the advantage of strong MIR molecular absorption features rather than direct detection of weak overtones with a near-infrared source.

%%%%%%%%%%%%%%%%%%%%%%%%%%%%%%%%%%%%%%%%%%%%%%%%%%%%%%%
\subsection{Frequency calibration of VIPA images}

The frequency information of the absorption spectrum was spread across the CCD according to the dispersion law described in reference \cite{xiao04b}, which accounts for VIPA thickness, tilt angle, and beam characteristics.  In practice it is difficult to assign frequencies directly from this dispersion law, and it is necessary to employ some form of direct image calibration by referencing known absorption spectral features \cite{cossel10} or directly imaging the frequency comb source \cite{diddams07}.  Calibration of the frequency information contained in the VIPA images was instead accomplished by launching a 10~GHz Ti:Saph frequency comb \cite{bartels08} and CW 795~nm reference laser into the VIPA spectrometer along with the upconverted signal (Figure~\ref{fig.setup}).  When combined with measurements of the Ti:Saph repetition rate $f_{\rm rep,TS}$ and absolute frequencies of the CW reference ($f_{795}$) and upconversion pump ($f_{1064}$) lasers, an absolutely calibrated frequency grid was correlated to the original upconverted VIPA image.  The position of a particular mode of the Ti:Saph comb on the VIPA image (Figure~\ref{fig.vipa}d) was assigned a MIR frequency value
\begin{equation}
	f_{\rm mode}=f_{795}+m\times{\rm  FSR}+n\times f_{\rm rep,TS}-f_{1064}
\end{equation}
where ${\rm FSR}$ was the free spectral range of the VIPA, $m$ was the number of free spectral ranges (vertical stripes) separating the Ti:Saph comb mode and 795~nm laser positions on the image, and $n$ was the number of Ti:Saph comb modes counting up from the bottom of each free spectral range.  Image positions between the Ti:Saph modes were assigned by linear interpolation between successive modes.  The precise value of the VIPA ${\rm FSR}$ depended upon the spectrometer tilt angle, and was calibrated from the VIPA images based on comparison with $f_{\rm rep,TS}$ over the entire image range.  The accuracy limit of $f_{\rm mode}$ was expected to be about 100~MHz  due to the wavelength meter measurements of the CW lasers.  In principle it would be possible to simplify frequency calibration by directly resolving a MIR comb with either an increased MIR comb $f_{\rm rep}$ or selective filtering of comb modes \cite{diddams07}  in combination with improved VIPA optical quality.

%%%%%%%%%%%%%%%%%%%%%%%%%%%%%%%%%%%%%%%%%%%%%%%%%%%%%%%
\subsection{Measurement procedure}

With transmitted light intensity $I$ from incident intensity $I_0$  through a sample length $L$, the Beer-Lambert Law predicts an absorption coefficient of
\begin{equation}
	\alpha=-\frac{1}{L}\ln\left(\frac{I}{I_0}\right).
\end{equation}

To account for camera count offsets, optical losses, stray light, and variation of the source spectrum, a full measurement consisted of pumping out the cell to record an upconverted normalization spectrum (Figure~\ref{fig.vipa}a), filling the cell and recording a VIPA image of the upconverted gas transmission spectrum (Figure~\ref{fig.vipa}b), and blocking the MIR beam before the upconversion crystal to obtain a background signal level (not shown).  Images of the reference CW lasers and Ti:Saph comb (Figure~\ref{fig.vipa}d) were recorded to produce the frequency calibration grid, and each 2-D VIPA image was reconstructed into a set of frequency and camera count pairs.
The absorption coefficient $\alpha$ for each frequency step was calculated as
\begin{equation}
	\alpha=-\frac{1}{L}\ln\left(\frac{C_{\rm tran}-C_{\rm bg}}{C_{\rm norm}-C_{\rm bg}}\right)
\end{equation}
where $C_{\rm tran}$, $C_{\rm norm}$, and $C_{\rm bg}$ are the counts in the transmission, normalization, and background images.

%%%%%%%%%%%%%%%%%%%%%%%%%%%%%%%%%%%%%%%%%%%%%%%%%%%%%%%%%%%%
\section{Absorption spectra}

\subsection{Low pressure ${\rm CH_4}$}
\label{sec.methane}

\begin{figure}
\begin{center}
\resizebox{.8\columnwidth}{!}{ \includegraphics{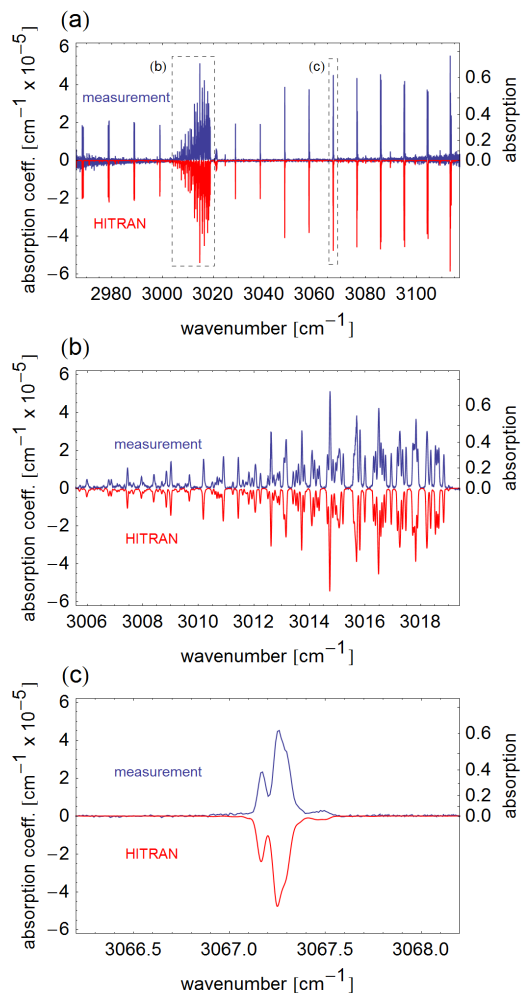}}
\end{center}
\caption{(a) Methane absorption signals (above axis) and predicted absorption spectrum from HITRAN database (below axis) in 210~m path length and 2.3~mTorr pressure.  (b) and (c) show narrow frequency regions of the the same data from (a).   Linewidths were limited by optical resolution of the VIPA spectrometer, and the HITRAN spectrum was convolved with the optical profile to compare directly to the measured spectrum.   The entire spectrum was recorded at a single MIR tuning position in a single 60~s measurement cycle (normalization, transmission, and background images).
}
\label{fig.methane}       
\end{figure}

The evacuated Herriott cell was filled with natural isotopic abundance methane gas (${\rm CH_4}$) at $2.3\pm.1~{\rm mTorr}$ total pressure. Figure~\ref{fig.methane} shows the measured absorption spectrum and a predicted spectrum calculated from the HITRAN database \cite{hitran08}.  The absorption spectrum was recorded at one MIR tuning position for the Raman soliton and DFG PPLN, using a single 5~s exposure for each of the transmission, normalization, and background images.  The 5~s exposure time was chosen to ensure no CCD pixels exceeded half of the the maximum well depth (77,000 electrons).  Total measurement cycle time of about 60~s was most limited by the pumping speed for gas exchange.   

At low pressure, linewidths were dominated by the optical resolution of the VIPA spectrometer (1.4~GHz or $0.048~{\rm cm^{-1}}$), so the HITRAN-predicted spectrum was convolved with the optical profile of the instrument to directly compare the spectra.  The absolute frequency scale in this data contains no adjustable parameters.  Comparison of the measured and predicted frequency scales shows agreement within about 150~MHz $(0.005~{\rm cm^{-1}})$,  comparable to the expected uncertainty in frequency calibration of $f_{\rm mode}$.  

Each VIPA order was recorded on the image (Figure~\ref {fig.vipa}) as a stripe with frequency increasing in the vertical direction.  All pixels in a stripe at the same vertical position had the same frequency value and were integrated horizontally to provide the count number in a single frequency step.  Each vertical pixel step on the image was separated by about 0.28~GHz ($0.009~{\rm cm^{-1}}$), which was smaller than the optical resolution of 1.4~GHz $(0.048~{\rm cm^{-1}})$.  In the highest intensity portion of the normalization image, the integrated electron counts in each vertical step  were  $C_{\rm norm}\approx65,000$.   The cooled CCD camera had low dark count and readout noise, so the dominant sources of noise were from shot noise in the electron count number ($\sqrt{C_{\rm norm}}$ and $\sqrt{C_{\rm tran}}$) and technical noise due to drift of the laser intensity or optical alignment. Decreased signal to noise at the edges of the spectrum was due to decreased spectral intensity of the MIR source in these regions.  

The best signal to noise ratio for a single frequency step would be expected with transmission that is neither very high nor very low.  At high absorption, the small remaining counts $C_{\rm tran}$ would be indistinguishable from $C_{\rm norm}$ noise, leading to a maximum detectable absorption coefficient of $\alpha_{\rm max,step}\approx2.6\times10^{-4}~{\rm cm^{-1}}$ for our experimental conditions.  At low absorption, the signal $C_{\rm norm}-C_{\rm tran}$ would be comparable to the total noise from $C_{\rm norm}$ and $C_{\rm tran}$, leading to a minimum detectable absorption coefficient of $\alpha_{\rm min,step}\approx2.6\times10^{-7}~{\rm cm^{-1}}$.  The range of $\alpha_{\rm min,step}$ to $\alpha_{\rm max,step}$ represents an estimate based on a single frequency step at the highest intensity portion of the image.  Changing the absorption path length $L$ would alter the maximum and minimum observable absorption coefficients while maintaining the same dynamic range, and spreading the absorption information over additional pixels would allow an increased dynamic range by avoiding the pixel well depth limit.

The limited VIPA optical resolution (1.4~GHz) resulted in a single spectral resolution element receiving contributions from about five 0.28~GHz frequency steps, so we would expect a statistical improvement of roughly $1/\sqrt{5}$ on the noise level for each spectral resolution element and a minimum detectable absorption coefficient of $\alpha_{\rm min,elem}\approx1.2\times10^{-7}~{\rm cm^{-1}}$ per element.  In the case of low pressure methane, the natural linewidth was narrower than the spectral resolution, so the minimum detectable absorption coefficient for a single peak should correspond to approximately  $\alpha_{\rm min,elem}$.  Additional technical noise or analysis in lower intensity portions of the VIPA image would decrease the detectable absorption range.

\begin{figure*}
\begin{center}
\resizebox{.7\linewidth}{!}{\includegraphics{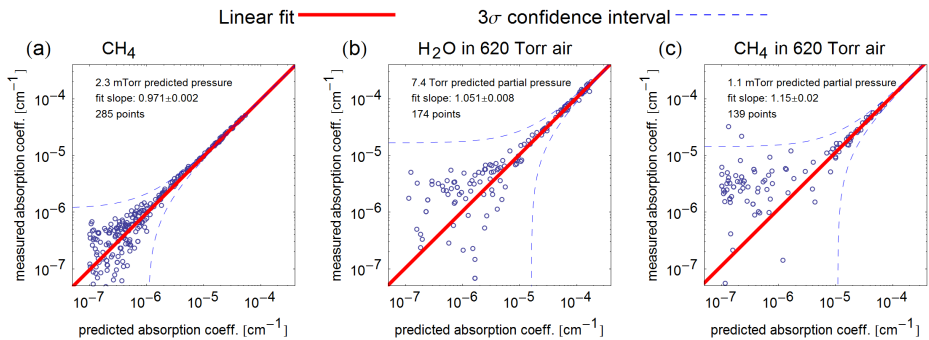}}
\end{center}
\caption{Measured versus predicted absorption coefficients for gas samples.  The linear fit slopes calibrate the measured pressure versus the predicted pressure, and the vertical intercept of the $3\sigma$ confidence interval shows the minimum measured absorption coefficient required to identify a single peak with 99.87\% confidence that it is not consistent with zero.  In all plots, the linear fits were constrained to pass through the origin.  The total number of measured peaks is also shown for each graph.}
\label{fig.hv}       
\end{figure*}

Another approach to determining the minimum detectable single peak absorption coefficient is based on the work of Hubaux and Vos \cite{hubaux70}.  
In this technique, measurements of a single line in a set of different sample concentrations are used to determine the minimum detectable concentration level that is not consistent with zero.  A modified version of this technique \cite{cossel10} uses the many available line strengths in a single recorded spectrum to make this determination, and we employed a similar method to analyze our system.  The predicted HITRAN spectrum for methane was prepared as a reference, including the optical profile convolution.  All locally prominent spectral features in the predicted spectrum with peak absorption coefficients $>10^{-7}~{\rm cm^{-1}}$ were selected as candidates.  Some of these features were not single lines, but contributions from several closely spaced lines.  

Absorption coefficient baseline offset errors due to drift of the laser intensity or optical alignment would be particularly noticeable at high or low absorption coefficients, but measurement of isolated spectral lines allowed a method to partially account for this drift by using the regions between absorption features.  Before fitting the absorption peak directly, a broader range of data around the peak (typically $4~{\rm cm^{-1}}$ to $6~{\rm cm^{-1}}$) was used to determine if a vertical offset was present in the measurement.  The  HITRAN model spectrum for this entire region was least squares fit to the data with a global amplitude adjustment to find the average baseline offset.   While holding the offset value fixed and then using only the data within the peak's narrow width, the HITRAN model was fit to the peak a second time to find the measured peak amplitude. The initially predicted pressure must be close to the actual pressure to limit significant changes in lineshape or position due to pressure broadening and shifting.  The baseline offset of the absorption coefficient varied nearly linearly across the entire spectrum by about $8\times10^{-7}~{\rm cm^{-1}}$, consistent with a spectral intensity drift of less than 2\% between normalization and transmission images.  

The measurement fit peak amplitude was plotted versus the peak amplitude of the original predicted HITRAN spectrum, and the process was repeated for the rest of the selected peaks.  Figure~\ref{fig.hv}a shows the resulting plot along with a linear fit.  The slope of the linear fit finds the measured methane pressure compared to the predicted pressure, and the uncertainty in the slope provides an uncertainty in the measured pressure.  The measured methane pressure was $2.233\pm.005~{\rm mTorr}$, consistent with the pressure gauge reading of  $2.3\pm.1~{\rm mTorr}$.  The $3\sigma$ confidence intervals that correspond to this fit are also shown.  The vertical intercept of the upper confidence interval can be interpreted as the minimum value of a peak's absorption coefficient that must be measured to have 99.87\% confidence that the measured feature is not consistent with zero.  The $1\sigma$ vertical intercept (not shown) of $\alpha_{\rm noise}\approx 4.4\times10^{-7}~{\rm cm^{-1}}$ corresponds to a peak height at the effective noise floor of the system (for a single spectral element).  This noise level approaches within a factor of four of the shot noise limited detection level $\alpha_{\rm min,elem}$, and is equivalent to less than 1\% fractional laser intensity noise.

The $3\sigma$ confidence intervals as presented here are for the measurement of a single peak within the spectrum.  In the context of a concentration measurement, the sensitivity can be improved by using many different peaks to average down the noise on the individual peak detections.  One implementation of the  multi-line detection advantage is to include a statistical improvement  in the Hubaux-Vos fit of $1/\sqrt{N}$, where $N$ is the number of spectral peaks within the measurement.  This approach is evident in \cite{cossel10} from the close proximity of the $3\sigma$ confidence intervals to the data spread, although the resulting $3\sigma$ detection limit on the vertical intercept is more difficult to interpret as a practical figure of merit.  Alternatively, comparison of the individual frequency step noise floor with the total integrated absorption in the measurement bandwidth \cite{adler10} yields a theoretical concentration limit taking into account both the multi-line advantage as well as the varying absorption strengths of all lines.  Using our measured system noise floor and bandwidth, this analysis indicates a minimum concentration limit of about 2~parts-per-billion (ppb) would be expected for methane in pure nitrogen at a pressure of one atmosphere. Additional measurements would be necessary to verify this.

%%%%%%%%%%%%%%%%%%%%%%%%%%%%%%%%%%%%%%%%%%%%%%%%%%%%
\subsection{Atmosphere}

\begin{figure*}
\begin{center}
\resizebox{.9\linewidth}{!}{\includegraphics{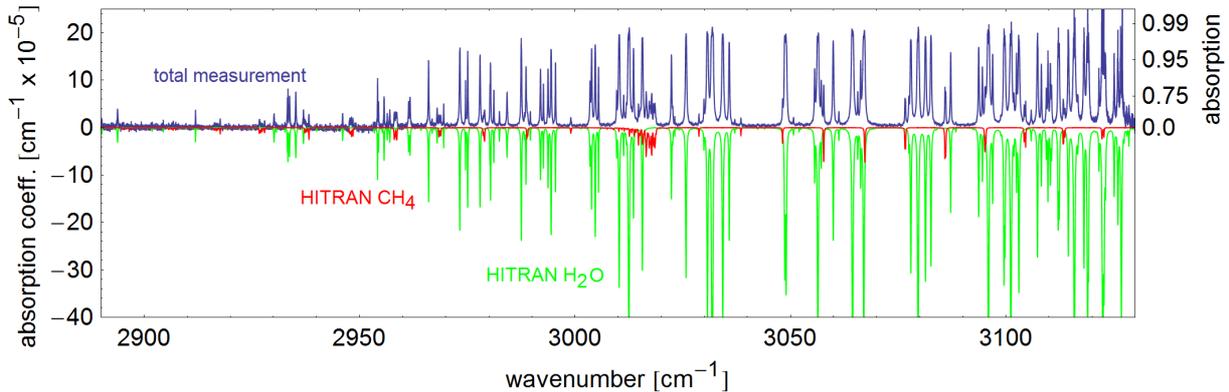}}
\end{center}
\caption{Absorption spectrum of laboratory air at 620~Torr.  The total measurement is shown above the axes, and HITRAN predicted spectra for methane and water vapor are shown below the axis.  Saturation effects are evident for the strongly absorbing water vapor peaks.}
\label{fig.air}
\end{figure*}

The gas cell was vented to atmosphere and a sample of background air from the laboratory filled the cell to total pressure 620~Torr.  The recorded spectrum is shown in Figure~\ref{fig.air} with HITRAN predicted spectra for ${\rm H_2O}$ (partial pressure 7.4~Torr) and ${\rm CH_4}$ (partial pressure 1.1~mTorr), where the pressures for ${\rm H_2O}$ and ${\rm CH_4}$ were initial estimates based on typical air composition and laboratory conditions.  No significant absorption features were observed for other species.   The entire span was again recorded at a single instrument position, with three 5~s recorded images and about 60~s total acquisition time.   Pressure broadening dominated the linewidths over the optical profile.  The measured spectrum began to show saturation as the absorption coefficient exceeded $\alpha_{sat}\approx 1.8\times10^{-4}~{\rm cm^{-1}}$.  As discussed in Section~\ref{sec.methane}, this level of saturation could be attributed to fractional noise or drift of the comb intensity of order 2\%.

To limit the effect of the saturated peaks, any frequency step with $\alpha>1.0\times10^{-4}~{\rm cm^{-1}}$ was removed from the analysis data before peak fitting.  The offset and peak fitting protocol was similar to that described in Section~\ref{sec.methane} and allowed independent variation of the amplitude of the water vapor and methane signals in the fitted model. The fit was still performed in regions where saturated points were removed based only on the wings of the peak (provided that at least half of the points within the linewidth were still present).  The measured versus predicted absorption coefficient comparisons with linear fits are shown in Figure~\ref{fig.hv}b-c with partial pressures of  $7.78\pm.06~{\rm Torr}$ for $\rm H_2O$ and $1.27\pm.02~{\rm mTorr}$ for $\rm CH_4$ that are within the expected range for the laboratory atmosphere conditions.  The methane concentration in a 620~Torr atmosphere can also be understood as $2.05\pm.03$~parts-per-million. 

 The minimum detectable absorption coefficients for both methane and water vapor were found to be ten times higher than the low pressure methane measurement, despite identical apparatus and technique.  This was also evident from the increased methane concentration uncertainty of 30~ppb as compared to the predicted performance of  about 2~ppb.  This noise floor increase was due to a combination of effects related to a ``real-world'' gas sample that includes overlapped spectral features, saturated peak absorptions, and additional optical scattering from airborne particles.

%%%%%%%%%%%%%%%%%%%%%%%%%%%%%%%%%%%%%%%%%%%%%%%%%%%%%
\subsection{Solvent mixture}

A tradeoff exists between the small wavelength range with high signal to noise of a tunable CW laser versus the wide coverage of the broadband  MIR/VIPA system with a decreased signal to noise for any single spectral line.  The tunable CW laser would retain a detection advantage when only  a small number of isolated lines are present, but as the density of lines and species increase or overlap, it would become challenging to determine a concentration with a single CW laser.  The broadband approach would benefit from additional lines within its span by using a collective fitting of all peaks to lower the detection limit for a gas species \cite{adler10}.  The difference between the CW laser and broadband laser becomes particularly apparent when measuring a continuous absorption spectra, and we show measurements of the solvents toluene (${\rm C_7H_8}$), gamma-butyrolactone (${\rm C_4H_6O_2}$), and isophorone (${\rm C_9H_{14}O}$) to illustrate this possibility.  These specific solvents were chosen because they have been identified as some of the compounds commonly present in the headspace around explosive materials \cite{lovestead10}.

A gas sample mixed from all three solvents was prepared to study concentration measurements of broad-featured gases.  The reference spectra for isophorone and gamma-butyrolactone were isolated from previous calibration with the MIR upconversion and VIPA spectrometer, and a high resolution quantitative absorption spectrum was used for toluene  \cite{chu99} .  The gas handling system was not designed to prepare this type of mixture precisely, so the partial pressures of each solvent were not well known.  In addition, contamination of the liquid solvents by exposure to atmosphere during mixing introduced an unknown ratio of other gases to the mixture.  Once expanded into the Herriott cell, the gas mixture had a total pressure of $232\pm1~{\rm mTorr.}$  

\begin{figure}
\resizebox{\columnwidth}{!}{ \includegraphics{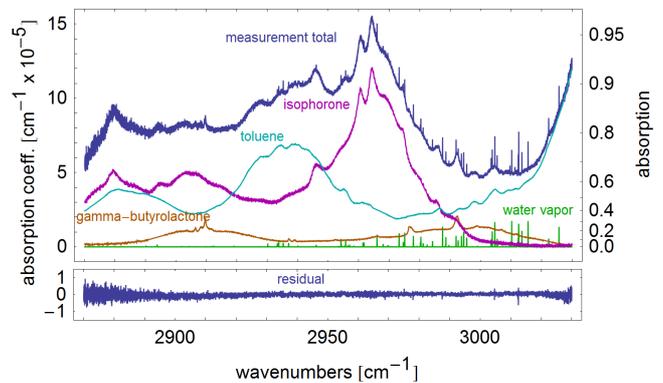}}
\caption{Measured total absorption of the solvent mixture (top).  The other spectra show the contributions of the constituents found by fitting the total spectrum with adjustable partial pressures.  Absorption coefficient fit residuals are shown below the main graph and have a standard deviation of about $9\times10^{-7}~{\rm cm^{-1}}$, within a factor of four of the shot noise limited minimum absorption coefficient for a single frequency step.
}
\label{fig.solvent}       
\end{figure}

The total absorption spectrum is shown in Figure~\ref{fig.solvent}.  The narrow peaks relative to the broadly varying background were compared to the HITRAN database and found to match ${\rm H_2O}$ that was unintentionally introduced during mixing.  No other narrow spectral features were found.  The partial pressures of the three solvents and water vapor were calculated by applying a least squares fit to match the total measured spectrum.  The independent absorption spectra of the constituent gases in the mixture are also shown in Figure~\ref{fig.solvent} for the fitted partial pressures of $142\pm6~{\rm mTorr}$ (${\rm H_2O}$), $18.6\pm.4~{\rm mTorr}$ (toluene), $57\pm6~{\rm mTorr}$ (isophorone), and $7.5\pm1.7~{\rm mTorr}$ (gamma-butyrolactone).  The largest sources of uncertainty in these pressures arose from the reference spectra used for calibration.  The total pressure based on the spectral fits of the constituent gases was $225\pm9~{\rm mTorr}$, which agrees with the total direct pressure measurement of $232\pm1~{\rm mTorr}$. The subtraction residuals (Figure~\ref{fig.solvent}) had a standard deviation of about $9\times10^{-7}~{\rm cm^{-1}}$ for the individual frequency steps (not resolution elements), a value that is within a factor of four of the expected shot limited single frequency step value $\alpha_{\rm min,step}\approx2.6\times10^{-7}~{\rm cm^{-1}}$ and corresponds to intensity noise levels of less than 2\% on the normalization comb spectrum.

The concentration precision in this demonstration was most limited by the non-ideal gas handling system for both the solvent mix and two of the reference spectra.  A more fundamental limitation when measuring broad absorption features is lack of opportunity to compensate for drift through the baseline correction methods used in the low pressure ${\rm CH_4}$ and air measurements, requiring a more stable system on both the DFG process and upconversion detection process.  Some of this drift could be eliminated by reducing the time interval between the normalization and absorption images with a faster pumping speed.

%%%%%%%%%%%%%%%%%%%%%%%%%%%%%%%%%%%%%%%%%%%%%%%%%%%%%%%
\section{Conclusion and outlook}

This system demonstrates the first high resolution spectroscopic application of femtosecond lasers in the mid-infrared without the complication of Fourier transform approaches using moving optical paths or multiple combs.  The $0.048~{\rm cm^{-1}}$ (1.4~GHz) resolution spanned up to $240~{\rm cm^{-1}}$ (7.2~THz) in a single measurement to yield up to 5000 distinct spectral elements.  The spectral elements exhibited a minimum detectable absorption coefficient of $\alpha_{\rm noise}\approx4.4\times10^{-7}~{\rm cm^{-1}}$, a value within a factor of four of the expected shot noise limited absorption coefficient.  This level of absorption coefficient noise could be expected to approach a concentration detection limit of 2~ppb for methane in pure nitrogen, and the present system has demonstrated methane concentration uncertainties at the level of 30~parts-per-billion in a more demanding direct atmospheric sample.

There are clear paths to improving the detection sensitivity of this system.  First, improved gas handling would increase measurement bandwidth by an order of magnitude and likely reduce intensity drift noise between absorption and normalization images.  Second, it is possible to increase the MIR power up to 25 times \cite{adler09} and the upconversion CW pump power more than 20 times to yield a 500-fold improvement in signal.  Taken together, these could lead to minimum absorption coefficients below $10^{-8}~{\rm cm^{-1}}$ per resolution element in 5~s acquisition times.  Finally, with appropriate change of the DFG and upconversion nonlinear crystals along with the CW upconversion pump laser, the system could span much further into the infrared using the same Yb:fiber femtosecond laser and near infrared VIPA detection.  The combined sensitivity, resolution, wavelength flexibility, bandwidth, and absence of moving parts make this upconversion femtosecond laser approach a viable and promising option for rapid detection of trace gases or separation of complex mixtures in a growing field of applications.

{\bf Acknowledgements}~~We would like to thank D. Richter for loaning the Herriot gas cell, A. Zolot for valuable discussions and assistance with sample preparation, and M. Hirano, Y. Kobayashi and I. Hartl for important contributions to the Yb:fiber frequency comb.  L. Nugent-Glandorf, T. Neely, E. Baumann, F. Giorgetta, and N. Newbury provided valuable comments and discussion.  T.A.J. acknowledges support from the National Research Council, and this work was funded by NIST and the the United States Department of Homeland Security's Science and Technology Directorate.

%%%%%%%%%%%%%%%%%%%%%%%%%%%%%%%%%%%%%%%%%%%

\end{document}